\def\b{\begin{equation}}
\def\e{\end{equation}}

\documentclass[prl,aps,twocolumn, showpacs, showkeys, floats]{revtex4}
\begin{document}

\title{Does Pressure Increase or Decrease Active Gravitational Mass
Density? }

\author{Abhas Mitra}

\email {Abhas.Mitra@mpi-hd.mpg.de}
\affiliation { MPI fur Kernphysik, Saupfercheckweg 1, D-69117
Heidelberg, Germany}

 \altaffiliation[Also at ]{Theoretical Astrophysics Section, Bhabha Atomic
Research Center, Mumbai-400085, India }



\date{\today}
\begin{abstract}
It is known that,  for a static fluid sphere, the General
Relativistic (GR) Effective  Mass Energy Density (EMD) appears to be
$(\rho + 3 p/c^2)$, where $\rho$ is the bare  mass density, $p$ is
the isotropic pressure and $c$ is the speed of light, from a purely
localized view point. But since there is no truly local definition
of ``gravitational field'', such a notion could actually be
misleading. On the other hand, by using the Tolman mass formula, we
point out that, from a global perspective, the Active Mass Energy
Density is  $\sqrt{g_{00}} (\rho + 3 p/c^2)$ and which is obviously
smaller than  $(\rho + 3p/c^2)$ because $g_{00} <1$.  Then we show
that the AGMD  eventually is $(\rho - 3p/c^2$), i.e., exactly
opposite to what is generally believed. We further identify the AGMD
to be proportional to the Ricci Scalar. By using this fundamental
and intersting property, we obtain the GR virial theorem in terms of
appropriate ``proper energies''.

\end{abstract}


\maketitle


Let us consider a static self-gravitating fluid sphere
described by the metric
\begin{equation}
ds^2 = g_{00}(r) dt^2 +  g_{rr} (r) dr^2 - r^2 (d\theta^2 + \sin^2
\theta d\phi^2)
\end{equation}
Here $\theta$ and $\phi$ are usual angular coordinates. The radial
coordinate $r$ is the luminosity distance or areal radius. While the
coordinate volume element is $dV = 4 \pi r^2 dr$ the proper volume
element however is $ d {\cal V} =  \sqrt{-g_{rr}}~dV $. Note that
since $-g_{rr} >1$, $d {\cal V}
> dV$. The energy momentum tensor of the fluid in mixed tensor form
is
 \begin{equation}
T^i_k = (p+\rho) u^i u_k +p g^i_k
\end{equation}
Here $g^i_k$ is the metric tensor, $u^i$ is the fluid 4-velocity,
$p$ is the isotropic pressure, $\rho$ is the total mass energy
density excluding any self-interaction and, hence, the {\em bare}
proper mass energy density. If $e$ is the total proper internal
energy density, $\rho$ includes $e$: $\rho = \rho_0 + e$ where
$\rho_0$ is the proper ``rest'' mass energy density not associated
with any active energy. Here we take $G=c=1$ unless $G$ and $c$
appear specifically.  The total mass energy appearing in Kepler's
law, i.e., the
 gravitational mass of the body is
\b
 M =M_g=  \int_0^R  \rho ~dV
 \e
 where $r=R$ defines the boundary of the {\em isolated} body having
 $p=0$ for $r >R$.
This total mass energy is also known  as {\sl Schwarzschild mass} of
the fluid.
 But it was shown by Tolman\cite{1} and Whittaker\cite{2} that there could be
 another definition of $M$.
For any stationary gravitational field, total four momentum of
matter plus gravitational field is conserved and independent of the
coordinate system used\cite{1,2,3}:
\begin{equation}
P^i = \int (T^{i0} + t^{i0}) ~dV
\end{equation}
where $t^{ik}$ is the energy momentum pseudo-tensor associated with
the gravitational field. Further,  the inertial mass (same as
gravitational mass),
 i.e, the time component
of the
  4-momentum of any given body in GR can be expressed
  as\cite{1,2,3}
 \begin{equation}
M = M_i= \int_0^\infty (T^0_0 -T^1_1 -T^2_2 -T^3_3) ~\sqrt {-g}~ d^3
x
 \end{equation}
 where
  $g = - r^4  g_{rr} g_{00}  \sin^2\theta$
is the determinant of the metric tensor $g_{ik}$ and $ d^3x =
dr~d\theta~d\phi$. Since \b T^1_1 = T^2_2 =T^3_3 = -p; \qquad T^0_0
= \rho \e
  it follows that\cite{1,2,3}
\begin{equation}
M = \int_0^\infty (\rho + 3p) \sqrt{-g_{00} g_{rr}} dV =
\int_0^\infty (\rho + 3p) \sqrt{g_{00}}  d{\cal V}
\end{equation}
When the body is {\em bound} and $p =\rho =0$ for $r \ge R$, then,
the foregoing integrals shrink to\cite{1,2,3}
\begin{equation}
M_i =  \int_0^R (\rho + 3p) \sqrt{g_{00}} ~ d{\cal V}
\end{equation}
This expression for $M=M_i$ widely gave rise to the idea that the
Active Gravitational Mass Density (AGMD) is $(\rho + 3p)$ and hence
pressure contributes positively to the AGMD.   Note that in
Newtonian gravity, the {\em bare} proper mass energy density is just
$ \rho_{Newton} = \rho_0$. But, in GR,  the same is $ \rho_{bare} =
\rho = \rho_0 + e $. Further, if the adiabatic index of the fluid is
$\gamma$,  in GR, the proper {\em bare} mass energy density is
$\rho_{bare} = \rho_0 + (\gamma -1)^{-1} p$. Thus the fact that
pressure would contribute positively
 to the effective mass energy is {\em already} taken into consideration
by the very definition of GR bare mass density. In fact, one sees
that pressure contributes practically linearly, almost on equal
footing to the inert rest mass density.

One may argue that since the RHS of the Einstein equation: $G^i_k =
(8 \pi G/c^4) T^i_k$ involves not only $\rho$, but also $p$, there
must be additional {\em positive} contribution of $p$ in AGMD over
and above what is hidden inside $\rho$. First, such a view is
imprecise because the Einstein equations are non-linear and
connected to each other in a subtle and complex manner which may
defy such a ``common sense''.   And if pressure would strictly
contribute to AGMD in a {\em positive} manner there must not be any
resistive action associated with pressure and, consequently, there
would be no hydrostatic equilibrium at all. Further, if  pressure
must contribute to $\rho$ on equal footing because both are
ingredients of energy-momentum tensor, it follows from Eq.(6), that
{\em pressure should contribute to AGMD in a negative manner}
because the signs of $T^0_0$ is {\em opposite} of those constituting
pressure! And this observation would be in tune with the notion that
in order that there is a hydrostatic balance, it is necessary that,
from global view point, pressure acts as a resistive element. In
general, one expects the AGMD to be determined as the {\em clothed}
mass density obtained  after accounting for all {\em global}
self-interactions of the {\em bare mass} which includes both rest
mass and pressure contributions. Since such global interaction
energies are necessarily negative, one does not expect the AGMD to
be larger than the bare mass density. Therefore, the interpretation
that \b \rho_{AGMD} = \rho + 3 p > \rho_{bare}\e is puzzling.

Note that $M_g=M_i$  is the total mass-energy perceived by distant
inertial observer $S_\infty$. Thus, the proper interpretation of
Eq.(8) is that while $\rho + 3p$ appears to be purely {\em locally}
defined total mass energy density, it need not represent the true
AGMD obtained after considering {\em global} negative self
-interactions. Thus we may call $(\rho + 3p)$ as the local Effective
Mass Density (EMD).  On the other hand,  only the unique global
inertial frame $S_\infty$  can perceive the true AGMD obtained after
accounting for global self-interactions. Indeed, a careful,
examination of Eq.(8) shows that, the actual AGMD appearing in
Tolman and Whittaker's mass is \b \rho_{AGMD} = (\rho + 3p)
~\sqrt{g_{00}} \e This may be further confirmed by noting that the
Poisson Equation, in this case\cite{4}, is \b \nabla^2 \sqrt{g_{00}}
= 4 \pi G ~\sqrt{g_{00}}~(\rho + 3p)\e Since Poisson equation should
indicate the true AGMD, it becomes clear that it is indeed
$\sqrt{g_{00}}~(\rho + 3p)$. Further, in the presence of mass
energy, $g_{00} <1$,  and hence it is less than $(\rho + 3p)$. Now,
let us see  whether  $\rho_{AGMD} < \rho_{bare}$ in
accordance with the fact that global self-gravitational energies,
either due to rest mass or internal energy or pressure or any other
source of mass energy is {\em negative}. In fact, we trivially
verify this in the following:

Note that the volume element $dV$ appearing in the Eq.(3) is {\em
not the proper element} $d{\cal V}$. To see the proper AGMD after
inclusion of self-energies, one must express $M$ in terms of proper
volume element:
\begin{equation}
M_g =\int _0^R \rho ~dV = \int_0^R {\rho\over \sqrt{-g_{rr}}}~
d{\cal V}
\end{equation}
 By the weak Principle of Equivalence (POE), the gravitational mass
 $M_g$ appearing in the foregoing equation must be equal to the
 ``inertial mass'' $M_i$, the temporal component of $P^i$ appearing in
 Eq.(8). Then, by comparing these two equations, we directly find that
 \b
 \rho_{AGMD} = (\rho +3p) \sqrt{g_{00}} = {\rho \over
 \sqrt{-g_{rr}}}\e
  Since in the presence of mass energy, $\sqrt{-g_{rr}} >1$, we
directly  verify from Eq.(13)  that
 \b
 \rho_{AGMD} < \rho
 \e
 because of  all pervasive negative self-interactions. It may be
 recalled here that
 the total {\em proper} energy content of
 the sphere, by excluding any negative self energy contribution\cite{4,5,6,7},
  i.e., the energy obtained by merely adding
 the {\em bare masses}
   is
\begin{equation}
M_{proper} = M_{bare}= \int_0^R \rho ~d{\cal V}
\end{equation}
Obviously, $M < M_{bare}$ because $dV < d {\cal V} $, and the
difference between the two constitutes the {\em conventional}
definition GR self- gravitational energy of the body \cite{4,5,6,7}
\begin{equation}
E_G =  M - M_{bare} = \int_0^R {(1- \sqrt{-g_{rr}})\over
\sqrt{-g_{rr}}} ~\rho~ d{\cal V}
\end{equation}
Since $\sqrt{-g_{rr}} >1$ in the presence of mass energy, $E_G$ is a
-ve quantity, as is expected. Thus irrespective of the subsequent
discussions, we have {\em already} shown in a trivial and direct
manner that $\rho_{AGMD} < \rho$ from a global perspective. Now we
proceed to find a more precise form of $\rho_{AGMD}$.

 Note that the $E_G$ defined by
Eq.(16) is the sum of appropriate local (proper) quantities somewhat
like the definition of $M_{bare}$ in Eq.(15); and is not defined
with respect to the unique inertial observer $S_\infty$\cite{4} 
who alone can define global energies from the perspective of energy
conservation. Note also that, in contrast, the gravitational mass
$M_g$ and inertial mass $M_i$ are indeed the mass-energy measured by
$S_\infty$. Accordingly, it was shown recently that the appropriate
value of the {\em global} self-gravitational energy as perceived by
the unique inertial observer sitting at spatial infinity
($S_\infty$) is different\cite{4}:
 \b {\tilde
E_{g}} =\int_0^R (\sqrt{-g_{00} g_{rr}} -1) ~\rho ~dV \e
The static GR scalar viral theorem, as perceived by the unique
inertial frame $S_\infty$ and obtained by merely demanding $M_g
\equiv M_i$\cite{4}
 is \b {\tilde E_{g}} = - \int_0^R 3 p ~\sqrt{g_{00}} ~
d{\cal V} \e
 This means that the effective  energy density of
self-interaction, as perceived by $S_\infty$ is simply
\begin{equation}
\rho_{self}^\infty  = - 3 p \sqrt{g_{00}}
\end{equation}
And since all {\em proper} energy densities are higher by the blue
shift factor of $1/\sqrt{g_{00}}$, the {\em proper density of global
self-interactions} is
\begin{equation}
\rho_{self} =  {\rho_{self}^\infty \over \sqrt{g_{00}}} = - 3 p
\end{equation}
Therefore, the net {\em proper clothed} mass density is \b
\rho_{clothed} = \rho_{bare} + \rho_{self} = \rho - 3p \e and which
is none other than the (negative of) {\em trace} of the energy
momentum tensor: \b \rho_{clothed} = -T^i_i = -T \e Since the AGMD
is the net clothed mass density after accounting for all positive
and negative sources of mass energy, we must have \b \rho_{AGMD} =
\rho_{clothed} = -T \e Had we used a different metric signature, we
would have found, $\rho_{AGMD} = T$. Let us rewrite: \b \rho_{AGMD} = \rho_0 + (\gamma -1)^{-1} p - 3p \e

Essentially pressure (also $\rho_0$ and $e$) remains submerged in
$\rho$ while this self energy is evaluated and {\em does not couple
separately to field}. In fact, the same is true in a Newtonian case
too where there is no ``weight'' with pressure. Even if the we would
have had $\rho_{AGMD} = \rho + 3p$ (in tune with prevalent notions),
it would have been incorrect to interpret the $3p$ term as directly
due to the ``weight'' of pressure. Had it been so, we would have had
just $\rho_{AGMD}=(\rho + p)$ rather than $\rho + 3p$! Of course, in such a
case, we would have had completely ignored the all self-interactions
in the determination of AGMD.

It would be more interesting to express this fundamental fact in
terms of the Ricci Scalar ${\cal R}$:

\b \rho_{AGMD} = {c^4 \over 8 \pi G} {\cal R} \e
 Conversely, we discover here
the fundamental physical significance of the Ricci Scalar, a very
important {\em invariant} of the problem: \b {\cal R} = {8 \pi
G\over c^4} \rho_{AGMD}\e And this observation is in strong
agreement with more general theorems on positivity of gravitational
mass\cite{8,9,10}. As a result of this study, we  obtained here the
expression for ``Active Gravitational Mass'' for the static
spherically symmetric fluid \b M= M_{AGM} =\int_0^R (\rho - 3p/c^2)
~d{\cal V} \e It may be remembered however that $M_g =M_i =
M_{AGM}$, and it would be naive to look for a separate ``pressure
contribution'' in AGM or in anything. What really contributes here
(negatively) is ``self-interaction'' (to ``bare mass'' $M_{bare}$).

For weak Newtonian gravity, we may directly verify Eq.(27) by
starting from the very definition of $M_i$ in Eq.(8). In the
Newtonian case, we have \b \sqrt{g_{00}} = 1 + \psi \e where $\psi
\ll 1 $ is the Newtonian gravitational potential. Following
Tolman\cite{1}, we may now split Eq.(8) into 4 terms by using
Eq.(28): \b M_i = \int \rho d{\cal V} + \int \rho \psi d{\cal V} +
3\int p d{\cal V} + 3 \int p \psi d{\cal V} \e From Eqs.(19) and
(28), note that  the strict expression for the Newtonian
self-gravitational energy is \b E_g^N = -\int 3 p d{\cal V} - 3 \int
p \psi d{\cal V} \e Now using Eq.(30) in (29), we find that \b M_i =
\int  \rho d{\cal V} + \int \rho \psi d{\cal V} - E_g^N \e

But the original definition of self-interaction energy is\cite{1} \b
E_g^N = {1\over 2} \int \psi ~\rho~d{\cal V} \e where the factor of
(1/2) comes because otherwise mass pairs would be counted twice. By
using Eq.(32) into (31), we have \b M_i = \int \rho d{\cal V} + 2
E_g^N -E_g^N  = \int \rho d {\cal V} + E_g^N \e i.e., the AGM is
indeed the clothed mass after accounting for the (negative)
self-interaction. Again using Eq.(30) into the foregoing Eq., we
obtain \b M_i = \int \rho d{\cal V} -3 \int  p d{\cal V} - 3 \int p
\psi d{\cal V}  \e Since in the Newtonian case, $p/c^2 \ll \rho$ and
$\psi \ll 1$, the last term in the foregoing equation drops out to
ensure that \b M_i = \int \rho d{\cal V} - 3\int  p d{\cal V} = \int
(\rho - 3p) ~d{\cal V} \e Tolman too obtained the same relation in a
somewhat less accurate way (see p.250 of Ref[1]). An absolutely
correct derivation of Eq.(27) is not obtainable in the Newtonian
case because of its inherent approximations. In contrast, the result
(27) was obtained in a strictly correct manner without Newtonian
approximations.

In the strict Newtonian case, one does not associate any mass
equivalence with either $p$ or $e$ or $E_g^N$.  Hence, the real
AGMD, $(\rho - 3p/c^2)$, does not appear in the Newtonian  Poisson
equation: \b \nabla^2 \psi = 4 \pi G \rho \e  Further, note that in
this case $\rho =\rho_0$ banishing all energies from the notion of
AGMD. However, in a Post Newtonian approximation, $(\rho - 3p/c^2)$
rather than ($\rho + 3p/c^2$) must replace $\rho$ in the foregoing
equation.

 The recently obtained GR virial theorem\cite{4} involves global quantities
 as measured by the unique inertial observer $S_\infty$. However,
 for astrophysical purpose, it would be more convenient to have a
 virial theorem which would involve conventional ``proper'' measure
 of self-gravitational interaction $E_g$ described in Eq.(16). Now
 we show that it is indeed possible to have this desirable
 astrophysical virial theorem by again equating $M_g$ of Eq. (12)
 with $M_{AGM}$ of Eq.(27):
 \b
 \int {\rho \over \sqrt{-g_{rr}}} ~d{\cal V} = \int (\rho - 3 p)
 ~d{\cal V}
 \e
 By recalling Eq.(16), one can easily manipulate the foregoing the above equality
 into
 \b
 E_G + 3\int p ~d{\cal V} =0
\e This is the desired GR virial theorem for a static spherically
symmetric fluid.

In summary, we found that

$\bullet$ Positive pressure of course adds to the locally defined
total mass energy density $\rho$ by lifting it from its inert value
$\rho_0$ to $\rho = \rho_0 + (\gamma -1)^{-1} p$. This however does
not mean that the AGMD is necessarily boosted by $p$.

$\bullet$ the AGMD $(\rho/\sqrt{-g_{rr}})$ is indeed lower than the
bare mass density $\rho$ because of all pervasive negative
self-interaction of all sources of gravity like ``rest mass
density'' and ``internal energy''. Pressure too contributes to this
interaction albeit via the  internal energy. The reduction of AGMD
in the presence of and internal energy and, hence, in the presence
of pressure may physically be seen as ``buoyancy'' acting upon  the
``bare mass'' immersed in a fluid.

$\bullet$ The question posed in the title of the paper is in
accordance with the current way of thinking in the community and it
is {\em ill posed} or at best naive. The more appropriate question
would have been whether pressure contributes to AGMD at all and

{\em ``Do self-interactions  contribute positively or negatively to
the Active Gravitational Mass Density?''}

As we found the $(-3p)$ term in the AGMD is not actually due to mere
pressure, on the other hand, it reflects the entire effect of
self-interactions. While calculating the self-interactions, both
$p$, $\rho_0$ and $e$ remain submerged within $\rho$ rather than
coupling independently to  $\rho$ or anything else. In the same
vein, $p$ contributes positively too to AGMD by through the $\rho$
term.

 $\bullet$ The result that AGMD decreases in a medium
because of interactions is in agreement with corresponding results
in nuclear and condensed matter physics.

$\bullet$ The present study concerns {\em isolated} objects where
$g_{00}$ could be uniform within the fluid only if it would be
unphysically considered that $p \to 0$. In such a case, $\nabla^2
\sqrt{g_{00}}  \to 0$ so that Poisson equation (11) would demand
both $\rho = 3 p = M \to 0$.

 The case is different in cosmology where one necessarily has an
 uniform $g_{00} =1$ despite having arbitrary pressure and EOS.
 However, the present discussion is valid for the segments of the
 universe.

 $\bullet$ We discovered here a fundamental result for a static
 spherical fluid: The Ricci Scalar is directly proportional to AGMD.
 This may have far reaching interesting consequences beyond the
 present study.

 $\bullet$ Finally we obtained a new and elegant form of GR virial
 theorem for a static spherical fluid which would be of potential
 importance in Relativistic Astrophysics.

\end{document}